# Numerical calculation of the transport coefficients in thermal plasmas

Ali Mahfouf, Pascal André and Géraldine Faure
LAEPT, Blaise Pascal University, 24 avenue des Landais,63177 Aubière Cedex, France

*Abstract-* We have performed a new efficient method to calculate numerically the transport coefficients at high temperature. The collision theory was treated to study singularities that occur when evaluating the collision cross section. The transport coefficients (viscosity, diffusion coefficient, thermal and electrical conductivity) depend strongly on nature of the interaction between the particles that form the plasma and that is why it is necessary to determine the interaction potential accurately.

## I. INTRODUCTION

THE properties of transport in plasmas and/or in gases at high temperature (>4000 k) play a key role in several applications, like the field of electric arc, plasma welding, plasma cutting and the field of aerospace problems for example atmospheric entry flights in which we need an accurate computation of the transport properties. The estimation of transport coefficients is very difficult to do experimentally at high temperature, thus the theoretical/numerical calculation is the preferred method. Two approaches solutions of the Boltzmann integro-differential equation have been proposed by Grad [1] and Chapman-Enskog [2] using the analysis of rigorous kinetic theory. The first method is called the Grad's moment method and second method Chapman-Enskog is the most used in the calculation of transport coefficients because of its effectiveness.

## II. COLLISION THEORY

The transport of matter, momentum and energy is due essentially to the collision between particles and the existing force fields. The study of interaction interparticular is crucial to understand the phenomena of transport. In the microscopic world the notion of collision between atoms or molecules is interpreted as an interaction between the forces associated with each of the interacting particles. In under our research we have considered that the plasma is in low-density, thus we can neglect the three-body interaction, and only elastic collisions have been taken into account.

### A. Binary Collision

We consider an elastic collision between two particles having mass $m$ and $M$, separated by distance $r$.

$$\chi(b,E) = \pi - 2b \int_{r_{min}}^{\infty} \frac{\frac{dr}{r^2}}{\sqrt{1 - \frac{v(r)}{E} - \frac{b^2}{r^2}}} \quad (1)$$

$b$, $r_{min}$ represent the impact parameter and the distance of closest approach, respectively, the angle of deflection $\chi$ is defined by the Eq.(1), $E$ is the total energy of the system of particles $m$ and $M$. $v(r)$ is the interaction potential between particles and is the most interested parameter for the study of the collision problem [1].

### B. Interaction Potential energy

There are many forms of interaction potentials energy to describe the same interaction, therefore it is hard to choose the good potential for a given interaction, however one can select it according the nature of the interaction force (repulsive or attractive). We can show in following examples some types of interactions that can be found in thermal plasma:

1- Charged ↔ Charged : this interaction is governed by the Coulomb interaction potential [3] but according the assumption of quasi-neutrality of the plasma [5], a shielding effect must be taken into account, consequently this binary interaction will be described by a Coulomb potential shielded at the Debye length $\lambda_D$ [5]:

$$V(r) = \pm V_0 \frac{\lambda_D}{r} \exp\left(\frac{-r}{\lambda_D}\right) \quad (2)$$

with: $V_0 = \frac{q_e^2}{4\pi\varepsilon_0 \lambda_D}$

A sign ($\pm$) depends on the sign of the particle charge (electron-ion).

2-Neutral ↔ Neutral: there are a lot of interaction potentials susceptible to describe this interaction, like Lennard-Jones potential, Morse potential, Buckingham Potential and others. We have opted for the Hulbert-Hirschfelder potential energy curve to describe this collision because it depends only on the spectroscopic constants [4].

3-Neutral ↔ Charged: the collision between the neutral and the ion can be studied by the same interaction as neutral-neutral.

### C. Collision cross section

In the case of the elastic collision, one define the total collision cross section $Q(E)$ as:

$$Q(E) = 2\pi \int_0^\pi (1-\cos(\chi))\sigma(\chi,E)\sin(\chi)d\chi \quad (3)$$

with $\sigma(\chi,E)$ is the differential scattering cross section:

$$\sigma(\chi,E) = \frac{b}{\sin(\chi)}\left|\frac{db}{d\chi}\right| \quad (4)$$

*D. Singularities problems*

We distinguish three main singularities when we evaluate numerically the collision cross section. One can remark in equation (4) that if (d$\chi$/db=0) then the differential cross section will be infinite ($\sigma(\chi,E) \to \infty$),this phenomenon is called the *Rainbow Scattering*.

The second singularity occurs when angle of deflection is a multiple of $\pi$ ($\chi = n\pi$) consequently sin($\chi$)=0 and the differential cross section will be infinite ($\sigma(\chi,E) \to \infty$) this singularity known as the *Glory Scattering*.

The third singularity that we are most interested is the *Orbiting* or *Spiraling Scattering*, physically, it means that when a particle collides with another one it can be that the two particles orbit each other, therefore the total cross section Q(E) may have many oscillations and the computing of Q(E) becomes very hard to do.

## III. TRANSPORT COEFFICIENTS

Gradients in concentration, velocity and temperature, cause a net transport or a net displacement of mass, momentum, and energy, respectively[5]. We can formulate mathematically this notion of transport as general form of a flux:

$$\vec{J} = -\alpha \vec{\nabla} \varphi \quad (5)$$

where $\vec{J}$ is the flux vector appropriate to quantity $\varphi$, and $\alpha$ is a proportionality constant known as the transport coefficient. The negative sign applies because the quantity is transported from regions of high concentration to regions of low concentration.

In the case of gradient in concentration, the transport coefficient is called the diffusion coefficient $D\ (m.^2s^{-1})$. Chapman-Enskog have formulated all transport coefficients, Eq.(6), Eq.(8) and Eq.(9), in term of collision integrals [1], [2], the diffusion coefficient was formulated in first approximation as follows:

$$D(T) = 2.628 \times 10^{-7} \left(\frac{T^3}{M}\right)^{\frac{1}{2}} \frac{1}{p\Omega^{(1,1)}(T)} \quad (6)$$

where *M, p,* and *T* are the molecular weight of the particle, the pressure in atmospheres, and Temperature in Kelvin, respectively. $\Omega^{(1,1)}(\overset{o}{A}^2)$ is the diffusion collision integral and is was formulated rigorously by Hirschfelder [1]:

$$\Omega^{\ell,s}(T) = \sqrt{\frac{kT}{2\pi m}} \int_0^\infty e^{-(\gamma)^2} (\gamma)^{2s+3} Q(E) d\gamma \quad (7)$$

with $\gamma$ =E/$k$T and Q(E) is seen in the previous section. $\ell$ ,s are the orders of approximation, $k$ is the Boltzmann constant and *m* is the mass of species.

The viscosity $\eta(kg.m^{-1}s^{-1})$ is the transport coefficient related to gradient in velocity expressed like this:

$$\eta(T) = 2.6693 \times 10^{-6} (MT)^{\frac{1}{2}} \frac{1}{\Omega^{(2,2)}(T)} \quad (8)$$

The flux correspondent is the viscous stress tensor [1].

In the case of gradient in temperature, the quantity transported is the energy which is related to heat flux vector Eq.(5) through the thermal conductivity $\kappa(J.m^{-1}.s^{-1}.K^{-1})$ expressed as:

$$\kappa(T) = 8.3227 \times 10^{-5} \left(\frac{T}{M}\right)^{\frac{1}{2}} \frac{1}{\Omega^{(2,2)}(T)} \quad (9)$$

## IV. NUMERICAL CALCULATION

In the previous section we remark that all transport coefficients

Eq.(6), Eq.(8) and Eq.(9) are expressed in function of so called collision integrals see the Eq.(7), and the integral collision is function of collision cross section see the Eq.(3) which is itself function of angle of deflection see the Eq.(1).

In summary, to compute the coefficients returns to evaluate efficiently three integrals which are included one into the other one. We have developed a numerical program which compute the three integrals in same time using the Clenshaw-Curtis quadrature [6], and we have also succeeded to eliminate the singularities that occurs principally in cross section integrals by introducing an adequate analytical formulates in the regions where occurs these singularities.

Our program has been tested, the results are compared with those of literature and they are in good agreement. We have computed the transport coefficient of Helium in different temperature see Table I and they are in good agreement with those of literature.

TABLE I
VARIATION OF TRANSPORT COEFFICIENT OF HELIUM WITH TEMPERATURE

| T(K) | $\eta$ (10$^{-6}$.PA.S) | $\kappa$ (10$^{-3}$.W. M$^{-1}$.K$^{-1}$) | D(10$^{-4}$.M$^2$.S$^{-1}$) |
|---|---|---|---|
| 100 | 9.410 | 73.308 | 0.263 |
| 400 | 24.277 | 189.109 | 2.793 |
| 1000 | 46.613 | 363.106 | 13.661 |
| 2000 | 77.922 | 606.99 | 46.349 |
| 4000 | 133.049 | 1036.420 | 160.755 |
| 5000 | 158.822 | 1237.182 | 241.290 |
| 10000 | 280.895 | 2188.0978 | 865.839 |